\documentclass[superscriptaddress,amsmath,amssymb,aps,prb,
                twocolumn,floatfix]{revtex4-2}
\usepackage{graphicx}% Include figure files
\usepackage{dcolumn} % Align table columns on decimal point
\usepackage{bm}      % bold math
\usepackage{hyperref}% add hypertext capabilities
\usepackage{cleveref}% allows multiple references at once
\usepackage{braket}
\usepackage{xcolor}

\begin{document}
\preprint{APS/123-QED}
\title{Efficiency of the hidden fermion determinant states Ansatz in the light of different complexity measures}
\author{Björn J.~Wurst}
\email{bjoern.wurst@rwth-aachen.de}
\affiliation{Institut f{\"u}r Theorie der Statistischen Physik, RWTH Aachen University and
  JARA---Fundamentals of Future Information Technology, 52056 Aachen, Germany}
\author{Dante M.~Kennes}
\affiliation{Institut f{\"u}r Theorie der Statistischen Physik, RWTH Aachen University and
  JARA---Fundamentals of Future Information Technology, 52056 Aachen, Germany}
 \affiliation{Max Planck Institute for the Structure and Dynamics of Matter, Center for Free Electron Laser Science, 22761 Hamburg, Germany}
\author{Jonas B.~Profe}
\affiliation{Institute for Theoretical Physics, Goethe University Frankfurt,
Max-von-Laue-Straße 1, D-60438 Frankfurt a.M., Germany}

\date{\today}
\begin{abstract}
    Finding reliable approximations to the quantum many-body problem is one of the central 
    challenges of modern physics. Elemental to this endeavor is the development of advanced numerical 
    techniques pushing the limits of what is tractable. One such recently proposed numerical
    technique are neural quantum states. This new type of wavefunction based Ansätze utilizes the expressivity
    of neural networks to tackle fundamentally challenging problems, such as the Mott transition.
    In this paper we aim to gauge the universalness of one representative of neural network Ansätze, the 
    hidden-fermion slater determinant approach. To this end, we study five different fermionic models 
    each displaying volume law scaling of the entanglement entropy. For these, we correlate the 
    effectiveness of the Ansatz with different complexity measures. Each measure indicates a 
    different complexity in the absence of which a conventional Ansatz becomes efficient. 
    We provide evidence that whenever one of the measures indicates proximity to a parameter 
    region in which a conventional approach would work reliable, the neural network approach 
    also works reliable and efficient. This highlights the great potential, but also challenges for 
    neural network approaches: Finding suitable points in theory space around which to construct
    the Ansatz in order to be able to efficiently treat models unsuitable for their current designs.
\end{abstract}
\maketitle
\section{Introduction}
The reliable prediction of material properties is central to designing new devices with taylored functionality~\cite{KITAEV20032, Castelvecchi_2023}. Elemental for this is the development of numerical techniques for treating the quantum many-body problem, the solution of which would allow to fully predict how materials behave. However, the full solution to this problem is impossible to obtain even for simple materials, and therefore approximate methods have been developed. Usually one is interested in the low temperature behavior of materials. To study this regime an effective low-energy model, derived via density functional theory~\cite{PhysRev.140.A1133, PhysRev.136.B864} or by other means~\cite{PhysRev.139.A796}, is constructed. This effective model is describing the material with a minimal number of degrees of freedom. Still these simplified models are often not exactly solvable and a number of numerical techniques, each with its strengths and weaknesses, have been developed.

These numerical approaches follow different lines of thought. A broad class are based on the Luttinger ward functional, i.e.~working on generating functionals. Examples of such approaches are dynamical mean-field theory~\cite{RevModPhys.68.13, RevModPhys.78.865, Rohringer_2018_routes, Vandelli_2022}, two particle self-consistent approach~\cite{Vilk_1997, Zantout_2019}, the Parquet approximation~\cite{PhysRevB.43.8044}, fluctuation exchange~\cite{PhysRevB.55.2122}, functional renormalization group~\cite{RevModPhys.84.299}, cluster perturbation theory~\cite{gros1993cluster,senechal2000spectral} and determinant quantum Monte-Carlo~\cite{PhysRevLett.90.136401, RevModPhys.73.33,Van_Houcke_2010}. Another large class aims to directly solve the stationary Schrödinger equation targeting the ground state (and a few excited ones) of the system. Examples for this class are matrix-product-state~(MPS) approaches such as density matrix renormalization group~(DMRG)~\cite{PhysRevLett.69.2863, Schollw_ck_2011, RevModPhys.93.045003,10.21468/SciPostPhysLectNotes.25}, wavefunction based variational Monte Carlo~\cite{gros1989physics, PhysRevB.71.241103, becca2017quantum}, the Gutzwiller-approximation~\cite{PhysRev.137.A1726, PhysRevB.85.035133, PhysRevB.96.195126, PhysRevMaterials.3.054605}, exact diagonalization~(ED)~\cite{PhysRevLett.72.1545} and neural quantum states (NQS)~\cite{carleo2017solving,lange2024architectures, MorenoHFDS2022, medvidovic2024neuralnetwork}. For the latter class the challenge is two-fold. First one has to design an algorithm allowing for an efficient ground state search:  in the case of ED for example the Lanzcos algorithm~\cite{Lanczos:1950zz}. This of course requires a way to store the ground state, posing the second challenge as the required memory scales exponentially with the system size. This storage problem is often solved by choosing a suitable predefined basis for the states representation~\cite{Schollw_ck_2011, 10.21468/SciPostPhysLectNotes.25} or a suitable compressing technique~\cite{PhysRevLett.132.056501}. For example, DMRG is relying on the entanglement entropy scaling in 1d gapped Hamiltonians which ensures that the formerly exponential cost for diagonalizing the Hamiltonian and storing the ground state is reduced to a polynomially one, enabling numerically exact studies for specific Hamiltonian types. 

Neural quantum states are a recently introduced approach of the second class~\cite{carleo2017solving,lange2024architectures, MorenoHFDS2022, medvidovic2024neuralnetwork}. Fueled by the fact that random neural network states chosen from a certain measure have volume law entanglement~\cite{PhysRevX.7.021021} a surge in numerical benchmark studies in spin~\cite{choo2019two, Ferrari2019Neural}, bosonic~\cite{mcbrian2019ground, stokes2023continuous, denis2024accurate} and fermionic systems~\cite{choo2020fermionic, MorenoFQ2020, MorenoHFDS2022, pescia2022neural, fore2023dilute, passetti2023can, gauvinndiaye2023mott, lange2023neural, d_schl2024neural, nys2024ab, kim2024, denis2024comment, dash2024efficiencyneuralquantumstates} have established this method as a versatile approach allowing to achieve high accuracy~\cite{wu2023variational, Chen2024}. However, while the numerical results are promising, a deeper understanding of what will and what will not be possible to solve in a faster than exponential scaling is lacking. Especially for fermionic systems there is currently a debate whether all physical ground states exhibiting volume law entanglement are treatable with exponential speedup or only some~\cite{passetti2023can, denis2024comment}. To understand this better, it is crucial to determine what limitations are inherited by closely related approaches such as DMRG and classical VMC trial states, especially with regards to expressivity of the trial state. To phrase it differently: Currently there is no a-priori understanding whether for a given Hamiltonian the ground state can be represented efficiently or not utilizing NQS.

A hint to the conditions under which a ground state is efficiently representable might be drawn from known complexity measures and their implications for related numerical techniques, such as the bipartite entanglement and the entropy of the Born-distribution:
Because NQS are usually able to represent matrix product states efficiently, they are also able to represent states with low entanglement
~\cite{chen2018, clark2018, deng2017, huang2021, sharir2022, wu2023, yang2024classicalneuralnetworksrepresent}. On the other hand, if the Born-distribution of the ground state has a low entropy, it indicates that the relevant Hilbert space size for the ground state search is reduced. Here the proximity to configuration interaction approaches enables NQS to efficiently solve the problem at hand~\cite{RevModPhys.32.300, DavidSherrill1999}. Furthermore, since fermionic NQS often build upon Slater determinants, the limit of weak-interactions should be representable from proximity to perturbative approaches. While all these limits are understood, the question remains how far out of these limits a neural quantum state Ansatz is still reliable and efficient.

\begin{figure}
\centering
\includegraphics[width=\linewidth]{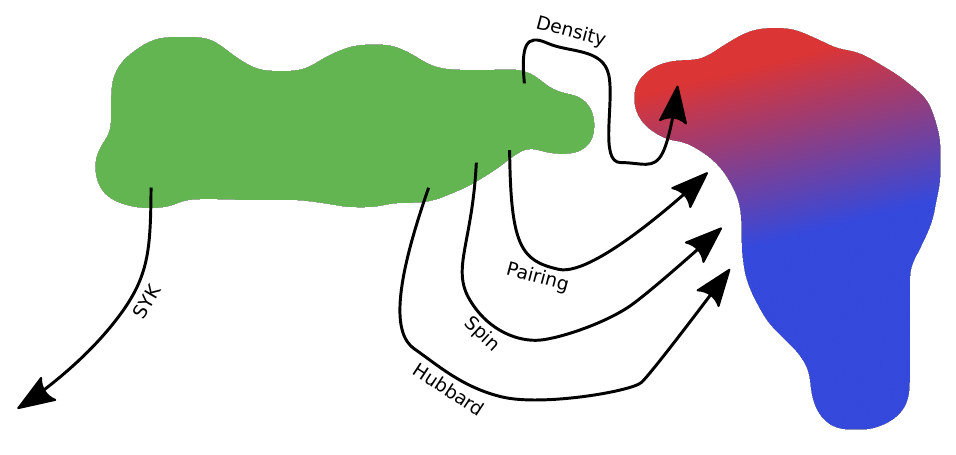}
\caption{Sketch of the regions in which we find that HFDS can efficiently solve the models. Green corresponds to uncorrelated states, blue to states with low entanglement and red highly fragmented states. In these areas solvability by NQSs with suitable architecture can be expected. The arrows are indicating the different models studied for which the arrow direction indicates an increase of the interaction strength. All models, except the SYK-model, are starting from a solvable region and approaching a solvable region in the large interaction limit. For all these models a reduction of the scaling of the number of parameters required compared to the dimension of the Hilbert-space will be observed. No such reduction is observed for the SYK-model.}\label{fig:summery}
\end{figure}

To delineate these limits, we consider a set of models with volume-law entanglement ground states, introduced in section~\ref{sec::models}. For each we track the number of parameters required to reach a certain accuracy in reference to the Hilbert space dimension. Further, we correlate our findings with both the presence of correlations and entropy of these systems, see section~\ref{sec::methods}, highlighting that performance is indeed best when expected from proximity to efficient solvability with another method. Our results, see section ~\ref{sec::res}, highlight the great versatility of neural network approaches to learn ground states of systems challenging for other numerical methods, while also outlining limitations. Further, we observe a strong correlation of the quality of our results with the quality of our Ansatz underlining the demand of model specific neural network state Ansätze in order to exploit underlying non-complexities.
These non-complexities are indicating closeness to represetable regions -- closeness to such a region implies a reduction of the complexity of the problem, as illustrated in Fig.~\ref{fig:summery}.

\section{Method}
\label{sec::methods}
\begin{figure}
    \centering
    \includegraphics[width=\linewidth]{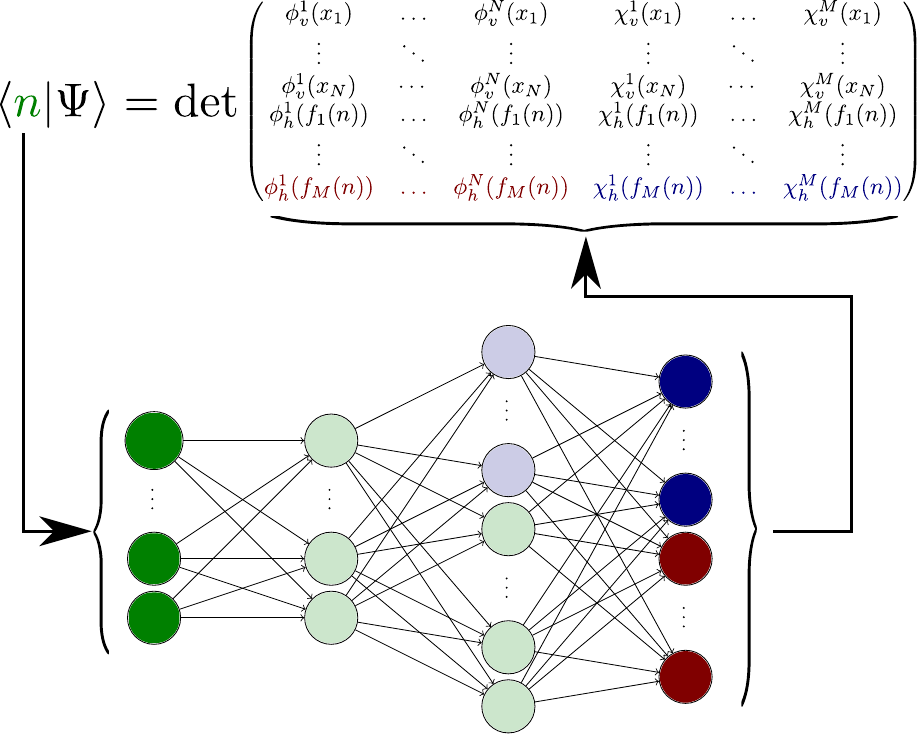}
    \caption{Sketch of the architecture of the HFDS state. The positions \(x\) are obtained from \(n\) by choosing an order by fixing the order. The signs corresponding to an exchange of particles is absorbed into the operators. The lower part of the matrix within the Slater-determinant is parameterized row wise by multi layer perceptions, in this sketch with two hidden layers.  The first half has the same dimension as the input, while the second half of the layers has the dimension of the input plus the numbers of hidden particles \(M\).}
    \label{fig:state_construction}
\end{figure}

\emph{Trial-state ---} For this paper we mainly want to explore NQS via the recently introduced hidden fermion determinant states (HFDS)~\cite{MorenoHFDS2022}. This approach showed promising results for different fermionic models~\cite{MorenoHFDS2022,gauvinndiaye2023mott}, nuclear matter~\cite{gnech2023distillingessentialelementsnuclear,PhysRevResearch.4.043178} as well as neutron star matter~\cite{fore2023dilute}. In HFDS the expressibility of an initial variational wavefunction is systematically improved by mapping an initial state \(\ket{\Psi_0}\) in an enlarged Hilbert-space \(\mathcal{H}^\prime\) containing \(N\) visible and \(M\) hidden particles into the target space \(\mathcal{H}\). The general form of the wave-function is given by
\begin{equation}
    \ket{\Psi}=\sum_n\ket{n}\braket{n,f(n)|\Psi_0}\,.
\end{equation}
By taking the initial state to be a Slater-determinant, a wavefunction of the form
\begin{equation}
    \Psi(n)=\left|\left|\begin{matrix}
        \phi_v(n)&\chi_v(n)\\
        \phi_h(f(n))&\chi_h(f(n))
    \end{matrix}\right|\right|\,.
\end{equation}
is obtained, where \(\phi_v(n)\) is a \(N\times N\) matrix and \(\chi_h(f(n))\) a \(M\times M\) matrix. The lower part of the matrix, i.e.~\(\phi_h\) and \(\chi_h\) are parameterized row wise by multi layer perceptrons as was done in Ref.~\cite{MorenoHFDS2022}. We utilize
\begin{equation}
    f(x+iy)=\operatorname{selu}(x)+i\operatorname{selu}(y)\,,
\end{equation}
as our activation function for neurons in the the hidden layers, i.e.~the scaled exponential linear unit~\cite{klambauer2017selfnormalizingneuralnetworks}
\begin{align*}
\operatorname{selu}(x)=\lambda\left\{\begin{array}{rc}
    x\,, & x\geq0 \\
    \alpha \mathrm{e}^{x} -\alpha\,,&\text{else} 
\end{array}\right.
\end{align*}
applied to the real and complex part separately. Here \(\alpha\approx1.67\) and \(\lambda\approx1.05\) are given by the default values used in the NetKet implementation~\cite{vicentini2022}. An exponential activation function is used for the output layer. The width of the hidden layers is given by the input dimension for the first half of the hidden layers and by the input dimension plus \(M\) for the second half. The architecture is sketched in Fig.~\ref{fig:state_construction}. Furthermore, we do also treat the visible sub sector as variational parameters opposed to using the natural orbitals corresponding to the lowest energies as was done in Ref.~\cite{gauvinndiaye2023mott}.

The necessity of building these more complex trial states in comparison to a plain neural network can be seen when considering certain non-interacting ground states, for example a fully connected random hopping model. Here no efficient representation can be found by feed forward neural networks~\cite{denis2024comment}, even though an efficient solution on classical computers is clearly possible.

The obtained wavefunctions are able to parameterize physically motivated trial-wavefunctions like Jastrow- or Gutzwiller-factors efficiently~\cite{MorenoHFDS2022}. Furthermore, they are closely related to neural network backflow states - as suggested by Ref.~\cite{liu2023unifying} they parameterize a similar variational manifold. Therefore, the results we present have implications for neural backflow states as well. 

\emph{Training ---} Our trial-states are trained on the overlap difference
\begin{equation}
    \delta \hat{O}=1-\hat{\mathcal{P}}_{\mathrm{tar}}\,,
\end{equation}
where \(\hat{\mathcal{P}}_{\mathrm{tar}}\) is the projector onto the target state~\cite{medvidovic2024neuralnetwork}. In the following, the ground-state of the system obtained with exact diagonalization. Training on this observable allowed to obtain larger fidelities with the target state in comparison to directly training on the Hamiltonian, especially in the case of small energy gaps to the low lying exited states. Because exact diagonalization already limits us to small system sizes and thus small Hilbert-spaces, we have calculated all observables by contractions over the full Hilbert-space instead of making use of the variational Monte Carlo method typically employed for this purpose. In doing so, we avoid an additional source for errors which would be introduced by the Monte-Carlo sampling.

For the training of the NQS we utilize the ADAM-optimizer~\cite{Kingma2014AdamAM} instead of the stochastic reconfiguration method more commonly employed for the optimization of NQS and variational states in general~\cite{PhysRevB.71.241103, MorenoHFDS2022, medvidovic2024neuralnetwork, gauvinndiaye2023mott, Chen2024}. For more details on this choice we refer to appendix~\ref{appendix:VarAnsatz+Opt}, where we elaborate on the reasons for this decision.

\emph{Observables ---} To estimate the quality of the fit provided by the neural quantum state, we use the expectation value of our loss-function, i.e.~\(\braket{\delta \hat{O}}\). This quantity is equal to one minus the fidelity of the variational state and the target state.

In order to estimate the complexity of the target state we use two quantities: The Shannon-entropy of the Born-distribution and the von-Neumann-entropy of the one particle reduced density matrix.
The Shannon-entropy of the Born-Distribution
\begin{equation}
    S_\psi=-\sum_n|\psi(n)|^2\log|\psi(n)|^2\,,
\end{equation}
is an upper bound to the bipartition entanglement entropy with an arbitrary bipartition compatible with the basis and serves as an indicator for the ground-state becoming sparse in the computational basis, i.e. fragmentation of the Hilbert space. This quantity therefore indicates that the state can be efficiently represented by configuration interaction (CI) methods. Because of the relation between the HFDS approach and CI methods shown in Ref.~\cite{MorenoHFDS2022}, one expects this also for the HFDS trial states. Furthermore, we want to mention a second case here: if the fragmentation is due to an operator diagonal in the computational basis, typical examples would be density-density-interactions or Ising-type interactions, the fragmentation can be captured by particular simple methods, like Gutzwiller-correlators. These are also closely related to the HFDS-approach.
A quantity which caries similar information, the inverse partition ratio, was recently also considered in Ref.~\cite{malyshev2024neuralquantumstatespeaked, li2023}.
As pointed out there are states with a large peak  in the electron density of the ground state wave functions, i.e. states having a large inverse partition ratio, hard to train. This is because it is hard to sample from the born distribution if it spans orders of magnitudes on the relevant space. Nonetheless, there exist improved sampling methods in this case making use of the closeness to CI~\cite{malyshev2024neuralquantumstatespeaked, li2023, cao2024visiontransformerneuralquantum}.
Low entanglement entropy as well as sparseness, i.e.~only a small fraction of the basis set of the Hilbert-space is relevant for the problem, should render the groundstate easy to represent.

The second quantity used to estimate the complexity of the target state is given by the von Neumann entropy of the \(N\) particle reduced density matrix \(\gamma_n\) (\(n\)-RDM)
\begin{equation}
    S(\gamma_n)=-\operatorname{Tr}\gamma_n\log \gamma_n\,.
\end{equation}
This quantity was analyzed in detail in Ref.~\cite{Carlen2016} and is a well known indicator for complexity for fermionic systems~\cite{PhysRevC.107.044318, PhysRevC.103.034325, PhysRevA.92.042326}. The most important property for our purpose is that \(S(\gamma_n)\) is bounded from below by states corresponding to Slater-determinants. This quantity therefore serves as an indicator of correlations - and consequently complexity - in the target state. For simplicity we will only consider the \(1\)-RDM. \(S(\gamma_1)\) is bounded from below by \(\log(N)\), where \(N\) corresponds again to the number of fermions in the system. This lower bound corresponds to the first \(N\) natural orbitals being fully occupied and the remaining ones being empty. In the opposite limit all natural orbitals are equally occupied.
Correlations are not only indicated by this quantity but also by the interaction strength. There are two main reasons for considering the 1-RDM instead: certain models have not only uncorrelated states in the non interacting limit, but also in the absence of a kinetic term: a large interaction strength does not necessarily imply strong correlations. This quantity is furthermore easier to handle when comparing different models, here correlations might be generated at different rates when raising the interaction strength.
Both of these quantities, i.e. the Shannon entropy of the Born-distribution and the von-Neumann entropy of the 1-RDM, are able to capture different non-complexities. We emphasize that also other non-complexities exist but for the reasons descried above we focus on these two.

\section{Models}
\label{sec::models}
\begin{figure*}[!thb]
    \centering
    \includegraphics[width=\linewidth]{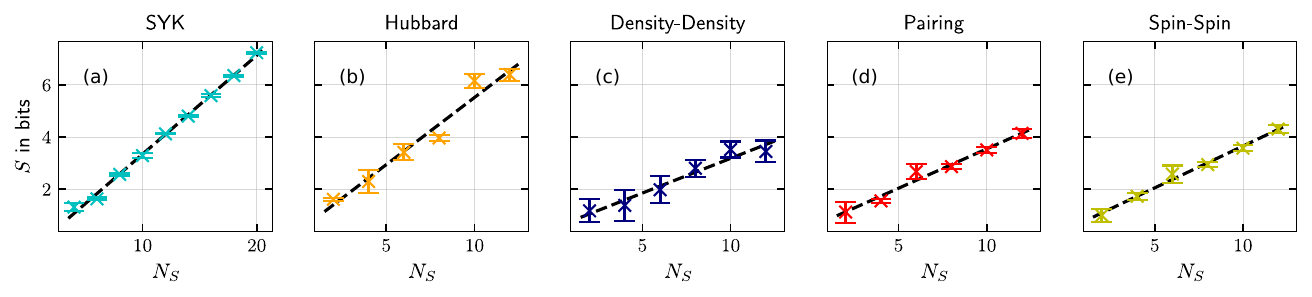}
    \caption{Bipartition entanglement entropies for the ground states of the various models considered here: (a) SYK-model, (b) random hopping Hubbard-Model, (c) density-density-interaction, (d) pairing-interaction and (e) spin-flip interaction. \(U=3\) was used. All models seem to posses volume law entanglement. The error-bars correspond to the standard error of the mean estimated via the sample variance. Each sample contains six realizations for each system size.}
    \label{fig:Entanglement}
\end{figure*}
To gauge the predictive and expressive power of new numerical approaches, it is instructive to consider different types of prototypical models. Ultimately, the aim is to obtain an understanding of what types of models can be efficiently solved and which ones are not efficiently solvable by understanding what the underlying mechanisms for the complexity reduction are. Here, we build on Ref.~\cite{MorenoHFDS2022} pointing out the relations of the used approach and known variational Jastrow based approaches as well as the known connection between MPSs and NQS~\cite{chen2018, clark2018, deng2017, huang2021, sharir2022, wu2023}.
To probe in what sense limitations of these closely related approaches are inherited, we consider SYK-like models, meaning models with fully disordered long range interactions, which are difficult for matrix product statesWe introduce five different models: a pure SYK model, a random hopping Hubbard model, similar to a model already studied in Ref.~\cite{gauvinndiaye2023mott}, a density-density interaction model, which is similar to a model already studied in Ref.~\cite{denis2024comment}, a spin-spin interaction model and lastly a pair-pair interaction model.
The five models are given by
\begin{align}
    \hat{H}^{\text{SYK}}=&\frac{J}{\sqrt{2N_S}}\sum_{ij}J_{ij}^{[2]}\hat{c}_i^\dagger\hat{c}_j\nonumber\\
    &+\frac{U}{\sqrt{2 N_S}^3}\sum_{ijkl}J^{[4]}_{ijkl}\hat{c}_i^\dagger \hat{c}_j^\dagger\hat{c}_l\hat{c}_k \label{eq:sykmod}\\
    \hat{H}^{\text{Hubbard}}=&\frac{J}{\sqrt{N_S}}\sum_{ij\sigma}J_{ij}^{[2]}\hat{c}_{i\sigma}^\dagger\hat{c}_{j\sigma}+U\sum_{i}\hat{n}_i^\uparrow\hat{n}_i^\downarrow\\
    \hat{H}^{\text{Density}}=&\frac{J}{\sqrt{N_S}}\sum_{i\leq j,\sigma}t_{ij}\hat{c}_{i\sigma}^\dagger\hat{c}_{j\sigma}+h.c.\\
    &+\frac{U}{4\sqrt{N_S}}\sum_{i\leq j}\sum_{\sigma\sigma^\prime}v_{ij}\hat{n}_{i}^\sigma \hat{n}_j^{\sigma^\prime}+h.c.\\
    \hat{H}^{\text{Pair}}=&\frac{J}{\sqrt{N_S}}\sum_{i\leq j,\sigma}t_{ij}\hat{c}_{i\sigma}^\dagger\hat{c}_{j\sigma}+h.c.\nonumber\\
    &+\frac{U}{\sqrt{N_S}}\sum_{i\leq j}v_{ij}\hat{c}_{i\uparrow}^\dagger\hat{c}_{i\downarrow}^\dagger\hat{c}_{j\downarrow}\hat{c}_{j\uparrow}+h.c.\\
    \hat{H}^{\text{Spin}}=&\frac{J}{\sqrt{N_S}}\sum_{i\leq j,\sigma}t_{ij}\hat{c}_{i\sigma}^\dagger\hat{c}_{j\sigma}+h.c.\nonumber\\&+\frac{U}{\sqrt{N_S}}\sum_{i\leq j}v_{ij}\hat{c}_{i\uparrow}^\dagger\hat{c}_{i\downarrow}\hat{c}_{j\downarrow}^\dagger\hat{c}_{j\uparrow}+h.c.
\end{align}
where the real and imaginary parts of \(t_{ij}\) and \(v_{ij}\) are drawn from normal distributions, i.e. \(\mathcal{N}(0,1)+i\mathcal{N}(0,1)\) and \(J_{ij}^{[2]}\), and \(J_{ijkl}^{[4]}\) are random unitary Gaussian variables full filling \(J_{ij}^{[2]}=J_{ji}^{[2]*}\) or \(J_{ijkl}^{[4]}=-J_{ijlk}^{[4]}=-J_{jikl}^{[4]}=J_{klij}^{[4]*}\) and \(\braket{|J_{ij}^{[2]}|^2}=1\) or \(\braket{|J_{ijkl}^{[4]}|^2}=1\) respectively. When not stated otherwise, we use \(J=0\) for the SYK-model and \(J=1\) for the remaining models. \(N_S\) stands for the number of sites of the model. 

From an entanglement entropy perspective, all five models follow a volume law scaling in \(N_S\) for the system sizes considered in Fig.~\ref{fig:Entanglement}. Nonetheless, they exhibit a large degree of structure; all the information necessary for their construction can be compressed down to at most \(\mathcal{O}(N_S^2)\) parameters, except for the SYK-model, where only a compression  into \(\mathcal{O}(N_S^4)\) parameters is possible. This also corresponds to a lower bound of the complexity: the Hamiltonian can usually be reconstructed from the groundstate \cite{Chertkov2018,Qi_2019, Hou_2020}. Therefore, both the variational manifold, due to the variational parameter number, as well as the set of ground states lie within a low dimensional sub manifold of the Hilbert space. Not a lot is known about the overlap of these two manifolds.

\section{Results}
\label{sec::res}
\begin{figure*}[!t]
    \centering
    \includegraphics[width=1.0\linewidth]{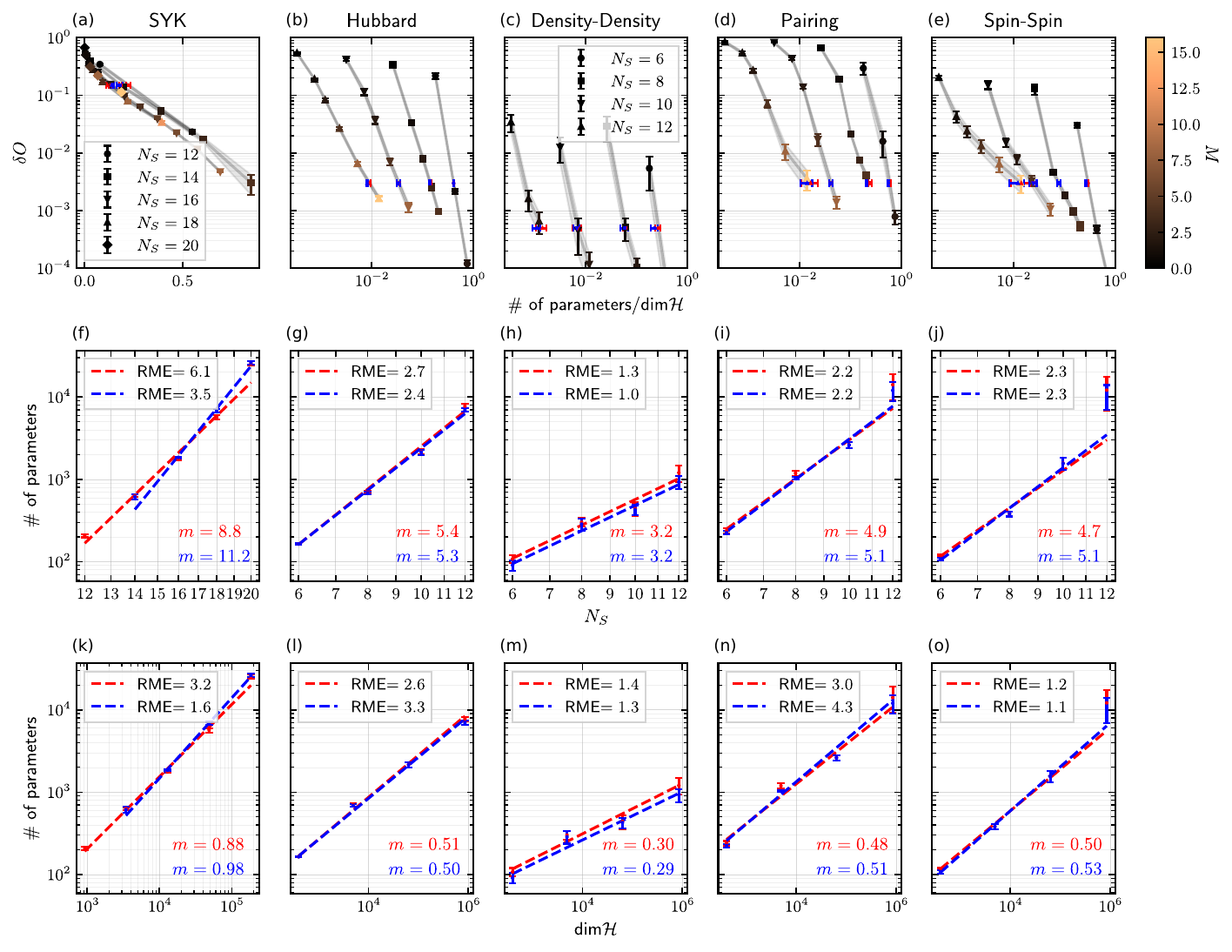}
    \caption{(a-e) Achieved errors for the various models and System sizes at \(U=3\): (a) \(N_S\) from \(N_S=12\) to \(N_S=20\) for the SYK-model and two hidden layers and (b-e) from \(N_S=6\) to \(N_S=12\) for the remaining models without hidden layers with the total magnetization constrains enforced (see also \cref{appendix:VarAnsatz+Opt}).
    The expressivity is controlled via the number of hidden particles indicated by the coloring of the data points.
 The order of the columns is the same as in fig.~\ref{fig:Entanglement}. The blue point corresponds to the globally interpolated estimates for the number of parameters for a fixed \(\delta O\) and the red points correspond to the locally interpolated estimates. Details can be found in the \cref{appendix:interpolation}. The color of the points encodes the number of hidden particles $M$.  In the second (f-j) and third row (k-o) we plot the interpolated estimates of the number of parameters to reach a certain accuracy and test two different fitting hypothesis, \(\propto N_S\,^m\) and \(\propto\operatorname{dim}\mathcal{H}^m\). In (f-j) we assume an algebraic scaling of the number of parameters with the respective \emph{system} size, while in (k-o) we assume an algebraic scaling of the number of parameters with the respective \emph{Hilbert}-space size (meaning exponential in system size). The quality of the fit is assessed by the root mean error (RME), for which we have rescaled the deviation by the error of the mean, which for most models is of the same quality between the two hypothesis.}
    \label{fig:huge}
\end{figure*}
To map out what is easy and what is hard to represent for the HFDS states, we calculate the overlap distance to the ground state as a function of the ratio of the number of parameters in the neural network and the Hilbert-space dimension. We plot these two quantities for all five models and different system sized in Fig.~\ref{fig:huge} (a-e). We observe two qualitatively different behaviors of the considered models: for the plain SYK model the number of parameters per dimension show a collapse of the different system sizes onto a single curve, whereas in contrast, for all other models the curves for different system sizes drift towards the lower-left - indicating that larger system sizes reach the same numerical error at a smaller relative number of parameters. The speed of the movement is found to be dependent on the specific model. These two different behavior patterns indicate different complexity scaling - while for the plain SYK model we find an exponential scaling, the movement observed for all other models indicates a deviation from a scaling trend. Whether or not this scaling becomes truly polynomial, stays exponential or shows some different sub-exponential behavior is not determinable by this data alone, as emphasised in the panels (f-o) of Fig.~\ref{fig:huge}. Both polynomial and exponential fits are compatible with the data. We verified this method of determining the scaling by performing the same analysis for an area law ground state, which then has a much smaller root-mean error (RME) for a polynomial scaling, see Appendix~\ref{appendix:area-law}.
These results highlight an important property of NQS when compared to other simpler variational trial states, e.g.~Slater-determinants with Jastrow correlators: they offer a controlled way of adding more parameters to the trial state and therefore to achieve the capacity necessary in order to achieve a given target error. The actual convergence can be prohibitively slow as indicated by the results for the SYK-model, making small errors in practice hard to achieve even for small system sizes.

While we do find a decreasing relative numbers of parameters required for reaching a certain error bound it should be noted that the absolute number of parameters still increases significantly, here the model with the density-density-interaction might be considered an exception. Reasons for this observation will be explained in more detail below. Furthermore, we find that the Hubbard-model, the model with the spin-spin-interaction and the model with the pairing interaction all require a similar number of parameters to reach a given error as can be seen in the second and third row of Fig.~\ref{fig:huge}. Here, generally a reduction in complexity had to be expected due to the reduction in entanglement entropy when compared to the SYK-model at similar dimension of the Hilbert space, as can be seen in Fig.~\ref{fig:Entanglement} (also compare \ref{app:ent_structure} for the Hubbard-model). Since the Hubbard-model shows a far greater entanglement entropy, while the other two models have an entanglement entropy similar to the density-density interaction model, as can be seen in Fig.~\ref{fig:Entanglement}, we conclude that the performance of the trial states can not be explained by only considering the entanglement entropy. Therefore, the pressing question arising is what determines the performance and can we assess it from inspecting the model?

\begin{figure}[!hbt]
    \centering
    \includegraphics[width=\linewidth]{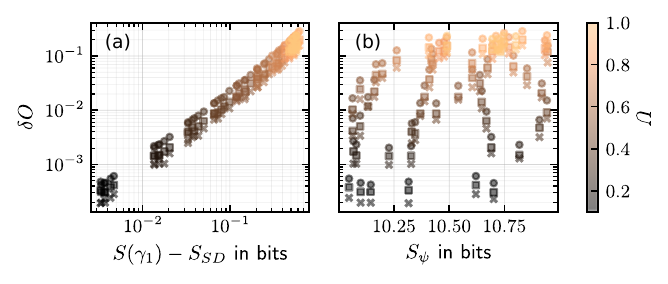}
    \caption{\(\delta O\) as a function of von-Neumann entroy (a) and entanglement entropy (b) for the \(14\) sites SYK-model for various combinations of kinetic and potential energy and seeds for the generation of the Hamiltonian. Here we fix \(J+U=1\). We visualize the results for $\{0,1,2\}$ hidden layers symbolized by \(\{\circ,\square,\times\}\) respectively. The trial wave function always contains one hidden particle. Independent of network depth, the error is highly correlated with the von-Neuman entropy of the \(1\)-RDM and rather uncorrelated with the entanglement entropy.}
    \label{fig:SYK-vanNeu}
\end{figure}

Two possible quantities to consider were already outlined above, namely the von-Neumann entropy of the 1-RDM and the Shannon entropy of the Born-distribution capturing the presence of correlations and fragmentation of the Hilbert-space respectively. We aim to explain observed features in terms of these two quantities.

Before discussing our results, we briefly mention what one would naively expect of the HFDS approach. As was shown in Ref.~\cite{MorenoHFDS2022}, with the HFDS approach we can represent ground states which are constructed by applying Jastrow-factors onto the Slater determinant basis. Thus allowing to represent weak-coupling states. Furthermore due to the proximity to CI like methods, solving the Hamiltonian in cases in which CI works well, namely when there is a Hilbert-space fragmentation, also works well. Keeping these arguments in mind we further investigate the central question considering $\braket{\delta \hat{O}}$ as a function  of the von-Neumann entropy of the 1-RDM and the Shannon entropy of the Born distribution for the SYK model and the fully connected Hubbard-Model.

In Fig.~\ref{fig:SYK-vanNeu}, we track the deviation for different numbers of hidden layers for different realizations of the SYK-model defined in Eq.~\eqref{eq:sykmod}: we observe a strong correlation between the increase of the deviation and the von-Neumann entropy of the \(1\)-RDM. This indicates that the correlations are not fully captured by the Ansatz. At the same time no strong correlation between \(S_\psi\) and the error is observed, as \(S_\psi\) stays approximately constant for all cases considered here. The observed success when representing the model in the weakly correlated case is not surprising, as it expected due to the Slater-determinant based construction of the trial state. An increase of the network's expressibility on the other hand via the depth is only leading to very minor improvements in achievable infidelity throughout the whole parameter range. Thereby indicating that the neural networks are struggling with the intricate correlation structure of the groundstate independent of the interaction strength. This implies that the state is not only hard to represent due to to the volume law scaling of the entanglement entropy (compare Appendix~\ref{appendix:area-law}), but also due to the correlated nature of the groundstate. A slightly different view on this is outlined in Appendix~\ref{appendix:sign-stucture}.

\begin{figure}[!hbt]
    \centering
    \includegraphics[width=\linewidth]{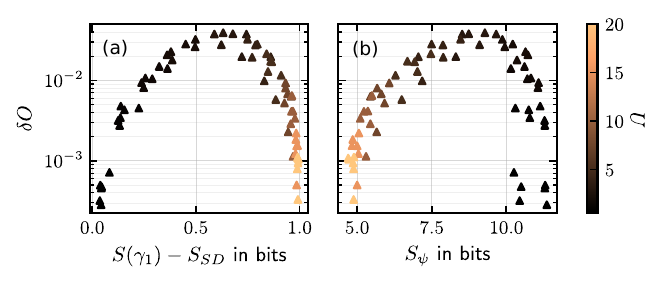}
    \caption{\(\delta O\) as a function of von-Neumann entroy  of the 1-RDM (a) and the entropy of the Born-distribution (b) for various interaction strengths for the fully connected Hubbard-Model with eight sites for \(M=1\). As can be seen the achieved error is small, if either the \(S(\gamma_1)\) or \(S_\psi\) are small. Here no hidden layers were used. The $t$ is fixed to one.}
    \label{fig:Hub-U}
\end{figure}

A somewhat different behavior can be observed in Fig.~\ref{fig:Hub-U}. While we find a similar trend, i.e.~a strong correlation between entanglement of the \(1\)-RDM and the error of the trained states for small \(U\), the behavior for large interactions is vastly different. 
At large interactions, we observe that the state becomes more sparse as states with double occupied sites are mostly projected out by the large interaction. This structure can be represented by the trial states considered here. 
The remaining part of the wavefunction only has support on a small sub-sector of the Hilbert space, leading to an easier to represent ground state. Explicitly, we observe a reduction of complexity explaining the better performance by proximity to CI-like solvability. While this implies a reduction in complexity, the exponential scaling of the problem remains, as even the spin model corresponding to the large \(U\) limit features an exponentially large Hilbert-space. An alternative interpretation of the decrease of the error at strong coupling is the following: we are now targeting a spin-liquid phase, of which certain types can be solved efficiently by Slater-determinant based states as for example done in Ref.~\cite{PhysRevB.88.060402}. Therefore, one can again explain the behavior by proximity to a perturbation theory around Slater determinants, exactly as in the weak coupling limit.
For the chosen system size the worst performance also corresponds to the point where the approximate capacity of the model given by the number of parameters is approximately equal to the size of the relevant Hilbert space estimated via \(S_\Psi\). Compare also Appendix~\ref{appendix:sign-stucture}.

For the density-density-interaction model we observed a stark reduction in complexity, see Fig.~\ref{fig:huge}. This can be explained from the principles introduced above: at the interaction strength tested the state does not feature strong correlations and already shows fragmentation introduced by the interaction term (see Appendix~\ref{appendix:large-U}). While the absence of correlations in the weak interaction limit is expected, this absence in the strong interaction limit is surprising at first glance. In the strong interaction limit (i.e.~neglecting the kinetic part) two cases have to be distinguished: either the groundstate is unique or it is two fold degenerate with the degeneracy protected by a \(\mathbb{Z}_2\) symmetry. In the first case the groundstate is a basis state and thus uncorrelated. In the second case the degeneracy is lifted by the kinetic term for finite \(U\) and prevents the complete projection onto exclusively singly occupied states, see Appendix~\ref{appendix:large-U}.
As a consequence, we observe no strong correlations in our numerical study of this model even in the large $U$ limit, see Appendix~\ref{appendix:large-U}, which explains the reduction of the achievable error when compared to the other models.

This highlights a crucial point: not all states showing volume-law entanglement are hard to represent, but can instead be represented by simple variational wavefunctions other than MPSs. Closeness to these solvable regions can often be understood and probed by simpler observables as for instance done here via the entanglement entropy of the \(1\)-RDM and of the Born-distribution for the density-density-interacting model. Such behavior can be understood as a strength of neural network based approaches: they are able to exploit various non complexities. Thereby, NQS can be build on already existing  and well known trial states, e.g Slater-determinants with Jastrow- or Gutzwiller correlators. Moreover, they can also exploit locality and are able to find efficient compression schemes without explicit knowledge about the structure of the target state. This makes them a less biased tools when compared to other trial states. What structure is exploitable is dependent on the trial state chosen indicating a need for physically motivated Ansätze~\cite{denis2024comment, PhysRevLett.122.226401}.

\begin{figure*}
  \centering
  \includegraphics[width=0.9\textwidth]{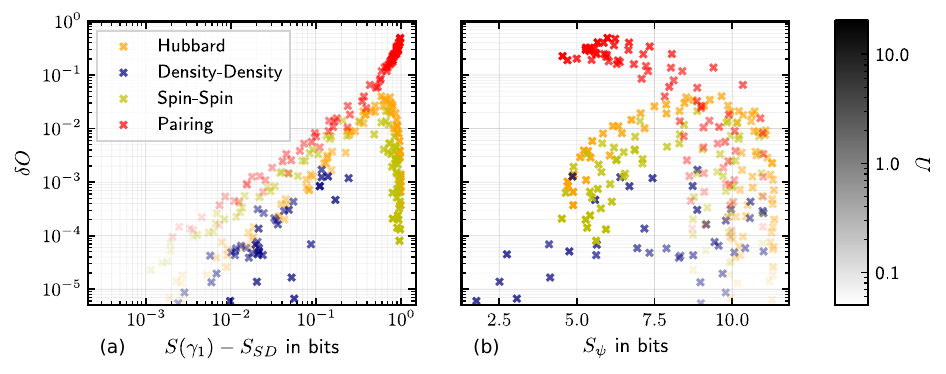}
    \caption{
        Results for the different entropies: in (a) the entropy of the \(1\)-RDM,  in (b) entropies of the probability distributions. The results were obtained with \(M=1\) employing no hidden layers. The system-size was chosen as \(N_S=8\) and six random realizations were considered. \(U\) ranges from \(0.1\) to \(20\), while the kinetic coupling was kept fixed at \(J=1\). The interaction strength is indicated by the alpha values of the data-points. Small alpha values, i.e. light points,  correspond to small interaction strengths, while large alpha values correspond to large interaction strengths. In subplot (a) it is clearly seen that the achievable error in the weak interaction limit is dependent on the localization of the electrons in the natural orbitals, indicating absence of strong correlation effects. Weak interaction in found on the left hand site of this plot. In subplot (b) we observe that the improvement in the strong interaction limit coincides with a reduction in complexity due to the electrons localizing in the computational basis. The data-points for strong interactions can be found on the left site of the plot.
    }
    \label{fig:Entropies}
\end{figure*}

In order to explore the connection between the ability of the NQS to represent the groundstate and the correlated nature of the target state we compare the different models with each other in Fig.~\ref{fig:Entropies}. Here, we considered a range of interaction strengths.
Especially for small interaction strengths a dependence between the presence of correlations and \(\delta O\) is observed. This strengthens our observations further -- the NQS considered here are not able to fully capture the correlation induced structure of the target state. For stronger interactions, we observe a deviation from this behavior in three cases. Firstly, for the density-density-model some data-points show smaller errors than expected when only considering \(S(\gamma_1)\). These points are characterized by a small entropy of the Born-distribution, compare Fig.~\ref{fig:Entropies} (b), and thus have a strong Hilbert space fragmentation. The second case in which a deviation is observed when comparing the spin-spin interaction model and the fully connected Hubbard model to the model featuring the pairing interaction. In the model featuring pairing interactions no increase of performance is observed for large interaction strengths, incontrast to the two other models. This deviation is understood as follows: In the large \(U\) limit the model featuring the pairing interaction is transitioning into a regime where most electrons are paired (see Appendix~\ref{fig:large-U}). This effect is notoriously hard to capture by Slater-determinant based approaches -- here a Pfaffian based approach should achieve better results~\cite{kim2024, PhysRevB.77.115112}. Nonetheless, a rapid decrease in error can be observed if either more hidden particles are considered (compare Fig.~\ref{fig:huge}) or more expressive neural networks with a single hidden layer are considered. Lastly, the density-density interaction model and the Hubbard model achieve larger fidelities for small interactions than the other two models when comparing points corresponding to similar values of \(S(\gamma_1)\). This feature is explained by the structure of the trial wavefunction (see Ref.~\cite{MorenoHFDS2022}): the excitations on a trial state with Jastrow-like prefactors obtained via a Laplace-expansion of the determinant strongly resemble states obtained by performing a few Lanczos-steps on the non-interacting wavefunction. Theerefore, the architecture is particularly suited to capture corrections to the noninteracting groundstates for these models.

\section{Conclusion and Outlook}
We have tested the ability of HFDS NQS to represent the groundstates of multiple prototypical models showing volume law entanglement. For these, we found two different behaviors: for the SYK-model the number of parameters necessary to achieve a given error is approximately proportional to the dimension of the Hilbert-space for all system sizes tested. This observation is in agreement with Ref.~\cite{passetti2023can} and implies an exponential scaling in system size of the parameters required to achieve a given error. The problem of representing the ground state of this model with NQS is therefore not due to the inability of multilayer perceptrons to learn a complex fermionic sign structure (see also Appendix~\ref{appendix:sign-stucture}) as speculated by Refs.~\cite{denis2024comment, gauvinndiaye2023mott}. Our results hint at a deeper problem when trying to represent groundstates of highly complex models like the SYK-model with neural-network based approaches instead.

For all other models we observe an improvement in the performance compared to the SYK-model, reproducing the results of Ref.~\cite{denis2024comment}. These models posses more structure than the SYK model, especially in the large interaction limit: the problem of representing the groundstate of the SYK-model is therefore not due to the volume law entanglement, but more generally due to the absence of exploitable structure for the variational wavefunction Ansatz. We have not been able to determine whether this extra structure is already sufficient to achieve sub-exponential scaling in the other cases.

Furthermore, these findings highlight a fundamental strength of NQS Ansätze: They have the ability to exploit various kinds of non complexities, effectively combining the strengths of other approaches, here combining Slater-Determinants with Jastrow or Gutzwiller correlators, CI and MPSs~\cite{MorenoHFDS2022}. Thereby, the NQS approaches utilize different sources of complexity reduction while other variational methods are often constructed around a single one. Whether the target state is close to such an inherited solvability region is often diagnosable by physical observables as demonstrated in this paper.  For example, the absence of correlations is indicated by a small entropy of the \(1\)-RDM, fragmentation by the entropy of the Born distribution and small entanglement by the bipartition entanglement entropy. While it is hard to understand how a neural network represents sufficiently complicated states, it is still possible to understand why an efficient representation can be found in many cases. This differentiates neural networks from interpretable methods like MPSs. When such non complexities are present, and the network is constructed such that they can be exploited, they allow for efficient compression schemes. However, when standard variational methods are not able to achieve good performance and no quantities usable for the compression are known, e.g.~for the SYK-model, no efficient compression scheme is found.
Furthermore, the closeness to other variational wavefunction approaches explains the good performance achieved for the density-density interaction model.

Our study suggests that while some volume law states can be represented efficiently with neural network based approaches, as indicated by Refs.~\cite{PhysRevX.7.021021, PhysRevLett.122.065301, PhysRevB.106.115138, passetti2023entanglement}, other physical volume law states seem still to be exponentially expensive and designing new approaches will be required to resolve these issues.  For example, to make progress on treating SYK like models, the most natural step is to extend known wavefunctions proposed in the literature~\cite{PhysRevResearch.3.023020, Bettaque_2024} with a neural network based Ansatz. For finding such architectures and furthermore understand inherited limitations a better understanding of when and why current approaches fail will be necessary.
Here our study has shed some light onto this boundary by studying the scaling of the parameters required for finding suitable solutions for various systems. Efficient solutions exist in case of weak correlations, strong fragmentation or local entanglement and all cases sufficiently close to such a limit.

\section*{Acknowledgements}
We thank Giacomo Passetti for fruitful discussions in the early stages of the project, further we thank Roser Valent\'i and Giuseppe Carleo for comments our mansucript.
Results were obtained using NetKet~\cite{vicentini2022, netket2:2019}, a Jax~\cite{jax2018github} based library for NQS.
Computations were performed with computing resources granted by RWTH Aachen University under project thes1588.
JBP, DMK acknowledge funding by the DFG under RTG 1995, within the Priority Program SPP 2244 ``2DMP'' --- 443274199.
JBP greatfully acknowledge support from the DFG through FOR 5249 (QUAST, Project No. 449872909, TP5).
DMK acknowledges support by the Max Planck-New York City Center for Nonequilibrium Quantum Phenomena.

\bibliography{bib}

\appendix
\section{Variational Ansatz and Optimization}\label{appendix:VarAnsatz+Opt}

\begin{figure}
    \centering
    \includegraphics[width=0.45\linewidth]{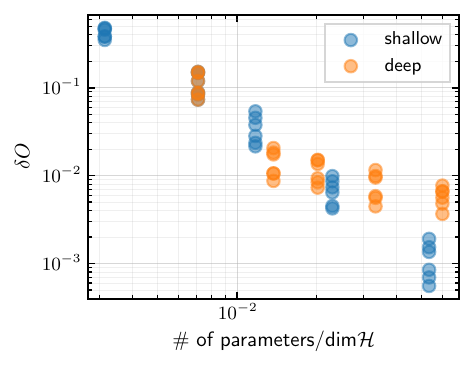}
    \includegraphics[width=0.45\linewidth]{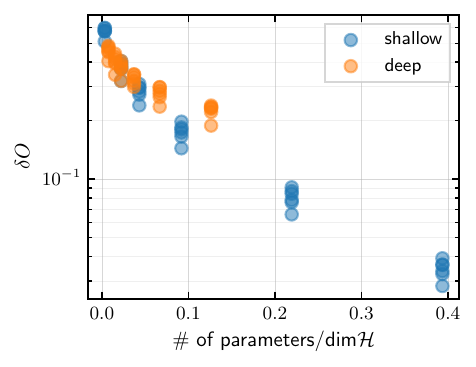}
    \caption{Comparision of the performance of deep and shallow networks at the same parameter number. On the left we show the results for the complex hopping Hubbard model with \(N_S=10\) and on the right for the SYK model with \(N_S=18\). For the former, we consider shallow networks with no hidden layers and various numbers of hidden particles compared to deep networks with only one hidden particles. For the SYK model, we consider two hidden particles for the deep network.
    In general, shallow networks are outperforming deep ones at the same number of parameters.}
    \label{fig:deep}
\end{figure}

\emph{Deep and Shallow Architecture ---} For many of the calculations shown in this paper shallow neural networks were employed, even though it is known that deep neural networks are able to outperform shallow networks in their ability to approximate functions efficiently in some cases. I.e.~for some functions an exponential increase in efficiency can be observed~\cite{pmlr-v49-eldan16, Lin2017}. Furthermore, various works imply an advantage of deep architectures over shallow architectures specifically for the problem of targeting physical wavefunctions~\cite{Gao2017, Carleo2018}\footnote{Here deep architecture means deep Boltzmann machines. Deep Boltzmann machines might lead to exponential increase of cost, even for only polynomial parameter numbers. The situation considered there is therefor different to feed forward neural networks, where an polynomial increase in parameters implies an polynomial increase in computation time for a evaluation of the network. Deep Boltzmann machines are in this regard similar to higher dimensional tensor networks.}. Therefore we want to address possible concerns here that our observations are due to considering shallow networks instead of deep ones.

When choosing our architectures we mainly had two things in mind: Firstly for the non SYK-model defined in eq.~\ref{eq:sykmod} with conserved total magnetization the number of parameters compared to the dimension of the Hilbert space would be large for small systems and deep networks. This leads to the ability to learn the wavefunction up to very high accuracy even for a single hidden particle.
Nonetheless, the ability to learn the target state can not be interpreted as the neural network finding efficient representations. The number of parameters of the trial state is already approximately equal to the dimension of the Hilbert space. In other words: it is not necessary to learn the structure of the target state. The neural networks already have enough capacity achieve high accuracies on noise.
Secondly, we did not observe a significant influence of the depth in the ability to learn the target states when keeping the parameter number fixed, see Fig.~\ref{fig:deep}. This observation is in line with Ref.~\cite{MorenoHFDS2022} where the authors also found that adding more hidden particles generally leads to lower expectation values for the energy than growing out the neural networks.

\begin{figure}
    \centering
    \includegraphics[width=0.48\linewidth]{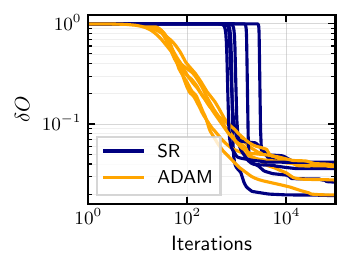}
    \includegraphics[width=0.48\linewidth]{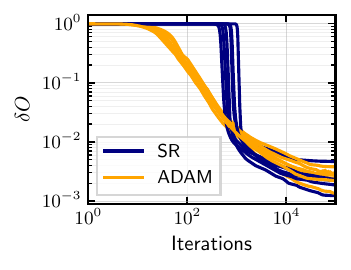}
    \caption{Comparison of the loss functions during the training of the random hopping Hubbard model with \(8\) sites and various seeds for the construction of the Hamiltonian for networks with no hidden layer and one/three hidden particle on the left/right. The learning rates are \(0.05\) and \(0.005\) respectively.}
    \label{fig:SR}
\end{figure}

\emph{Optimization Algorithm ---} For our optimization we have used the Adam algorithm~\cite{Kingma2014AdamAM} instead of the typically employed stochastic reconfiguration algorithm (SR) for the training of NQS. This does not seem to have an influence on the achievable model performance as can be seen in Fig.~\ref{fig:SR}.

Because we are we not able to employ the kernel trick proposed in Refs.~\cite{Rende2023ASL,Chen2024}, due to contracting over the whole Hilbert space instead of making use of variational Monte-Carlo for the calculation of the quantities necessary, SR becomes very expensive per step in comparison to the ADAM-algorithm for large models. Even when needing fewer steps to converge, ADAM is still cheaper in terms of computational resources.
Furthermore, the optimization via SR did sometimes get stuck at \(\delta O\sim0\). Which is most likely an effect of the spectral properties of the Projector.

Lastly we have observed that the quantum geometric tensor is ill behaved during some steps of the optimization, especially for the SYK-models. Thus we need to apply large regularization corrections to the quantum geometric tensor leading to sub optimal performance of this method. The low rank behavior of the quantum geometric tensor in certain cases was also studied in Ref.~\cite{dash2024efficiencyneuralquantumstates}.

Because of these reasons we concluded that the ADAM-optimizer is a better fit for our use-case.

\begin{figure}
    \centering
    \includegraphics[width=0.5\linewidth]{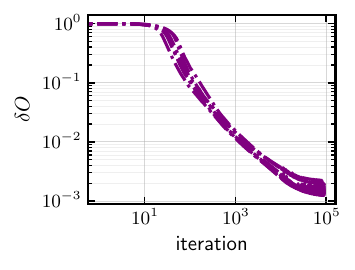}
    \caption{Training curves for the data points used for the Hubbard model in \cref{fig:huge} with \(N_S=12\) and \(M=16\). A flattening out of the training curves can be observed at the end of the training.}
    \label{fig:convergence}
\end{figure}

\emph{Convergence ---} For larger systems with many parameters, fully converging the training curves is becoming increasingly expensive. For this reason the training was only performed until the training process was flattening out, see Fig.~\ref{fig:convergence}. Here the training is transitioning into a regime where a small improvement in training accuracy would require an exponentially fast increasing number of training steps.

\section{Interpolation}\label{appendix:interpolation}
The data in Fig.~\ref{fig:huge} was interpolated (and extrapolated) in order to estimate the number of parameters needed to achieve a given accuracy for various system sizes. Here, the target error was chosen as small as possible without relying on extrapolation of our data to predict the number of parameters necessary for the error chosen. For the purpose of estimating the number of parameters necessary to reach a certain accuracy bound two methods were employed: a piece-wise and a global interpolation.

\emph{Piecewise interpolation ---} For the piecewise interpolation we estimated the mean \(\delta O\) as well as the error of the mean for each data point first and obtained the estimates for the number of parameters as well as the errors via a simple interpolation process between neighboring points. Here we used an algebraic fit, i.e. \(Ax^{-m}\) for all models except the SYK-model~\ref{eq:sykmod}, where we have observed a better agreement for the function \(A\mathrm{e}^{-mx}\), where \(x\) corresponds to the number of parameters.

A problem of this method is that errors due to convergence problems for large models are not considered as these are hard to train. The introduced bias towards larger parameter numbers for trial wave functions of large systems results in a bias towards the exponential scaling hypothesis. This problem should at least be partially be cured by the global interpolation.

\emph{Global interpolation ---} For the models discussed in the main text we have observed a good agreement of our data for a given system size to our fitting functions if one excludes the \(M=0\) point, i.e.~the pure Slater-determinants. Therefore, we additionally interpolated the data points globally for a given system size and a given seed used for the generation of the matrix elements appearing in the Hamiltonian. This allows us again to estimate the number of parameters given \(\delta O\). The reported estimate now corresponds to the mean of these predictions with the error bars estimated by the sample variance.

This approach is not as responsive to small errors due to convergence problems in the largest models. Still a good agreements between both methods can be observed for our data points. Therefore, we conclude that the increase of the computational cost observed above is not due to not fully converging out the largest systems.

 Because \(M=0\) and \(M=1\) are the only two data-points relevant for the smallest system with pure density-density-interaction, we kept the \(M=0\) point in this case. A similar problem occurred for the smallest system size for the SYK-model. Here we decided to skip this system size for this model for the global interpolation as we were able to achieve results for the same number of system sizes as for the other models considered.
 
 The difference in the fitting function for the global fit is necessary due to data-constrains. For the piece-wise interpolation corrections are in general still small (see also Fig.~\ref{fig:scaling1DSYK}). Our observed scaling for the SYK-model opposed to all of the other models observed in Fig.~\ref{fig:huge} is therefore not an artifact of differences in the interpolation method.

\section{Large \(U\)-Behaviour}\label{appendix:large-U}
\begin{figure}[htbp]
    \centering
    \includegraphics[width=\linewidth]{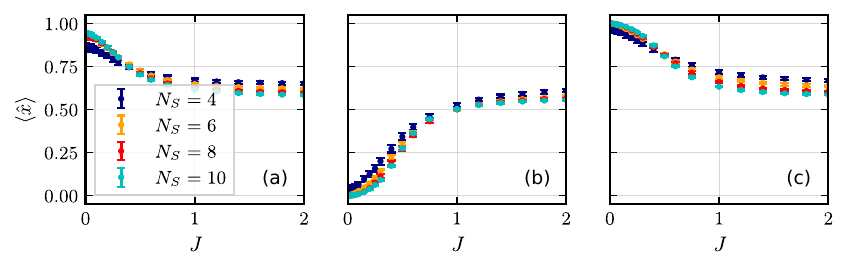}
    \caption{Expectation value of \(\hat{x}\) for the groundstate of the model with the density-density-interaction (a), spin-spin interaction (b), pairing interaction (c) averaged over various realizations. The scale of the interaction was fixed at \(U=1\). For the spin-spin model double occupations are completely suppressed, while for the  pairing based model we see that single occupations are suppressed. For the density-density model the behavior is more complicated due to the possible \(\mathbb{Z}_2\) symmetry.}\label{fig:large-U}
\end{figure}
 
In order to gain a better understanding of the behavior of the models featuring a density-density interaction, a spin-spin interaction and a pairing interaction in the limit of large coupling strengths, we calculate
\begin{equation}
    \hat{x}=\frac{1}{N_S}\sum_i 1+2\hat{n}_{i\uparrow}\hat{n}_{i\downarrow} - \hat{n}_{i\downarrow}-\hat{n}_{i\uparrow} \equiv \frac{1}{N_S}\sum_i \hat{x}_i
\end{equation}
for various system sizes and ratios \(J/U\). The results are visualized in Fig.~\ref{fig:large-U}. The \(\hat{x}_i\) is defined in such a way that it returns zero for single occupied sites and and one for doubly occupied ones. We observe in the model with the spin-spin interaction double occupations are suppressed for large interactions. The model becomes effectively a spin model.

A similar behavior can be observed for the model with the pairing interaction; here single occupations are suppressed. After performing a particle-hole transformation of the spin-down species, the model can be mapped back to the spin-spin model (in the absence of a kinetic term). Therefore, the model is transitioning into a regime with where all electrons are paired and single occupied sites are projected out. 

\begin{figure}[hbtp]\label{fig:DD_S}
    \centering
    \includegraphics[width=\linewidth]{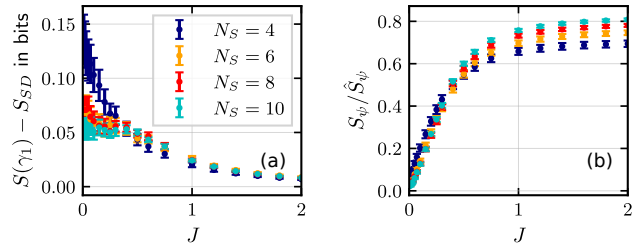}
    \caption{Correlations measured by \(S(\gamma_1)\) (a) and fragmentation measured by \(S_\psi\) divided by the maximal possible value (b) for the groundstate of the density-density-interaction model for various system sizes averaged over multiple realizations. Correlation stay small for all system sizes considered here, even though they are not tending to zero. This indicates a degenerate groundstate in the \(J\rightarrow0\) limit. For the strongly interacting case strong fragmentation can be observed.}
\end{figure}

For the density-density interaction model the expectation value of \(\hat{x}\) is not converging towards \(1\) for \(J\rightarrow 0\) for all system sizes considered here. This can be traced to the interacting Hamiltonian being already diagonal in the computational basis and we find that at least for some of the realizations the ground state is degenerate. The corresponding symmetry-group here is \(\mathbb{Z}_2\): This symmetry is originating from a computational basis state mapping either onto itself or a different computational basis state under a global spin flip. In the latter case we observe \(\braket{\hat{x}}\neq 1\) in the former \(\braket{\hat{x}}=1\). The degeneracy is lifted by the kinetic term. In Fig.\ref{fig:DD_S} (a) it can be seen that the degenerate groundstate appears for all system sizes considered. Subplot (b) shows that the target space is increasingly peaked in the strong interaction limit as would be expected. While the support of the target space seems to be growing with the dimension of the Hilbert space even in the strongly interacting case in the limit \(J=0\) this behavior brakes down. While seemingly still leading to an exponential grows of the relevant sub-sector the prefactor becomes arbitrarily small in this limit, indicating the (maximally) two dimensional space spanned by the ground states in this limit.

Furthermore, as indicated by \(S(\gamma_1)\) in subplot (a) of Fig.\ref{fig:DD_S}, the groundstate of this model is never strongly correlated -- \(S(\gamma_1)\) stays far below the maximal possible value. This renders these states simpler to learn for Slater-determinant based approaches than expected from considering the fragmentation.

These effects are explaining the ability to represent the groundstate of the density-density interaction model with fewer parameters compared to all other models. The state has exploitable structure, i.e.~it is close to a gaussian state and fragmented.

\section{Application to Area-Law States}\label{appendix:area-law}
\begin{figure}[!htbp]
    \centering
    \includegraphics[width=0.5\linewidth]{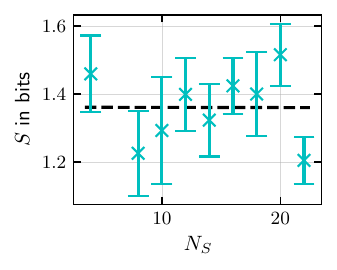}
    \caption{Entanglement entropy for various system sizes for the model defined in eq.~\ref{eq:1DSYK} with \(J=0\). Even for these small system sizes the entanglement entropy does not seem to be dependent on the system size, indicating area-law scaling of the entanglement entropy for this model.}\label{fig:1Dentropy}
\end{figure}
\begin{figure}[!bp]
    \centering
    \includegraphics[width=\linewidth]{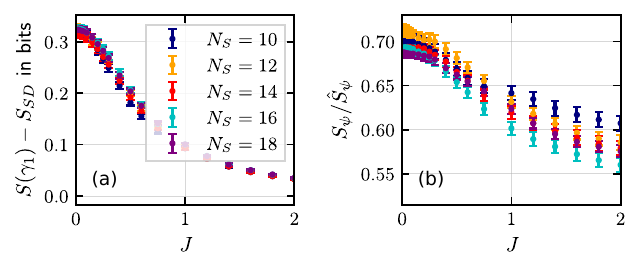}
    \caption{Correlations measured by the von-Neumann entropy of the 1-RDM (a) and the entropy of the Born-distribution (b) for various system sizes and various kinetic energy strengths given by \(J\). The interaction scaled was kept fixed at \(U=1\). In what follows we will be concerned with the case \(J=0\). In this case \(S_\psi\) indicates a broad born distribution as can be seen in subplot (b). Also \(S(\gamma_1)\) is indicating a correlated groundstate.}\label{fig:complexity1DSYK}
\end{figure}

\begin{figure*}[!h]
    \begin{minipage}{0.5\textwidth}
    \includegraphics[width=\linewidth]{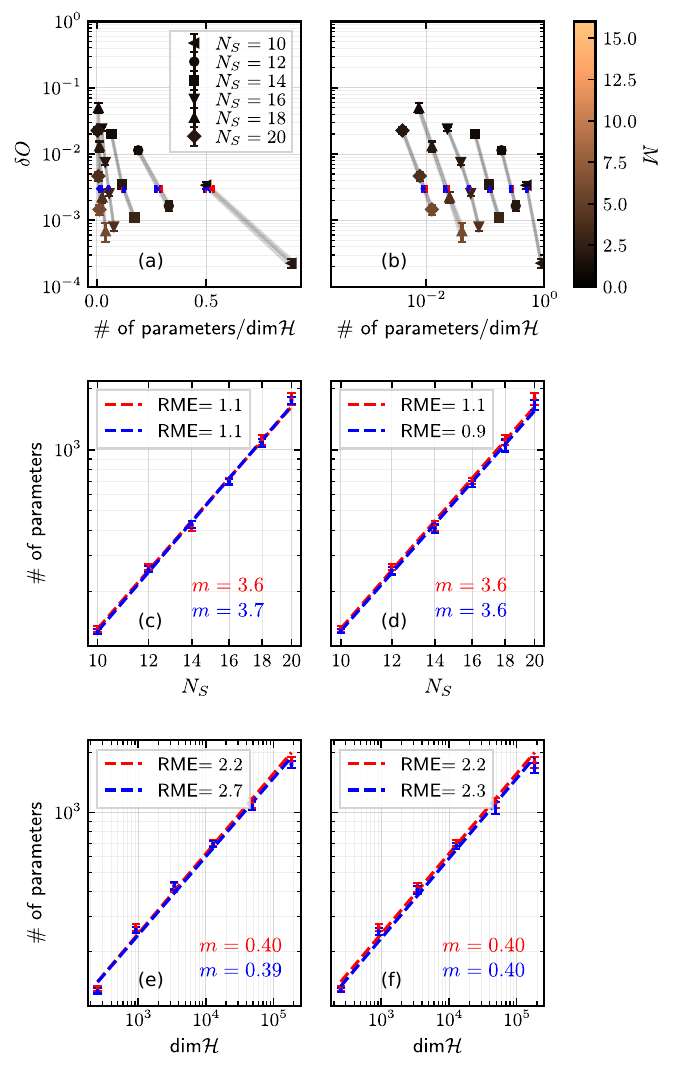}
    \end{minipage}%
    \begin{minipage}{0.5\textwidth}
        \caption{Estimates for the scaling of the number of parameters with the system size for model Eq.~\ref{eq:1DSYK}. For each system size 30 realizations of the model were used. Here no hidden layers were used. This choice was taken due to the necessity of considering models with fewer parameters in comparison to the fully connected SYK-model. For the left column the exponential hypothesis was used for the interpolation and for the right column the algebraic one was used (see Appendix~\ref{appendix:interpolation}). While the algebraic hypothesis provided a better fit for the data, the final results (c-f) show now strong dependence on the method chosen. Here the  The estimates for the scaling of the parameters can be found in sublots (c-f), first row with the polynomial scaling hypothesis and in the second row with the exponential scaling hypothesis. Here the polynomial is clearly favored over the exponential one. The RME is expected to take the value of one assuming a Gaussian distribution of the estimated means around their actual values. The deviation of the data point for the largest system in (c-f) can again be explained by the difficulty of training these larger models.}\label{fig:scaling1DSYK}
    \end{minipage}
\end{figure*}
This appendix serves the purpose of testing our method on one dimesional area-law states~\cite{Hastings_2007}. If we were not able to observe a reduction in complexity in this case, it would indicate a problem with our analysis.
To this end we consider
\begin{equation}
    \hat{H}=\frac{J}{2}\sum_{(i,j)\in S_2}J_{ij}^{[2]}\hat{c}_i^\dagger\hat{c}_j+\frac{U}{\sqrt{8}^3}\sum_{(i,j,k,l)\in S_4}J^{[4]}_{ijkl}\hat{c}_i^\dagger \hat{c}_j^\dagger\hat{c}_l\hat{c}_k\,,\label{eq:1DSYK}
\end{equation}
where \(J^{[2]}\) and \(J^{[4]}\) are sampled from the same distribution as they were for Eq.~\ref{eq:sykmod}. \(S_2\) is the set of tuples with indices at most one site apart, i.e. on-site and neighboring sites, and \(S_4\) is the set of quadruples where the elements are at most four sites apart. Results for the entanglement entropy can be found in Fig.~\ref{fig:1Dentropy} and the results for the degree of correlations and sparsity of the wavefunction are shown in Fig.~\ref{fig:complexity1DSYK}.
In what follows we will be concerned with the case \(J=0\). The target state is supported by a large subset of our computational basis, contains correlations and shows a small entanglement entropy that stays constant in system size, i.e. the groundstate of the system is a one-dimensional area-law state.

Due to these properties of the model we are able to use it to test the influence of low entanglement on the representability of ground states within the HFDS approach -- neither of the other two complexities implies that the state is particular easy to learn.

The scaling of the number of parameters with the system size is shown in Fig.~\ref{fig:scaling1DSYK}. Here we have chosen to show the results for the exponential interpolation and the algebraic interpolation. While the fit for the algebraic methods seems to be working better, the final result for the scaling of the number of parameters with the system size is largely independent of this choice.

Furthermore, beyond the reduction of the scaling in comparison to all models (except the density-density interaction model) a strong tendency towards polynomial scaling can be observed. This is indicated by the root mean error of the polynomial fit. From the reduction of computational cost, we conclude that the increase of the computational cost is not due to the method chosen for the estimation of parameters, but due to the complexity of the target states. In other words, Area-law scaling is an exploitable non complexity in one dimension. In order to make other states learnable other non complexities are necessary. In this light the trend towards polynomial scaling can be expected and small entanglement entropy can be added to the set of exploitable quantities.

While a scaling with exponent \(m\sim 3.6\) seems very expensive when compared to tensor network methods, we still want to highlight that in order to make this wavefunction extensive/size-consistent one has to choose \(M\propto N_S\), already implying \(m=3\) asymptotically. Still, this does lead to an increase of error in system size, as the error on different subsystems also needs to be reduced in order to keep the total error constant, implying \(m>3\).

\section{Phase structure of the Target State}\label{appendix:sign-stucture}
It was already observed that the ability of NQS to find suitable approximations for the target state is in practice often hindered by their ability to obtain sufficiently good approximations for the sign- or phase-structure of the target state~\cite{Westerhout2020}. To link this observation back to our observed behavior, we try to learn the sign structure of the ground state in a Slater-determinant basis. For this purpose we will train the trial state
\begin{equation}\label{eq:phase}
    \braket{n|\Psi_{var}}=\frac{\braket{n|\Phi}}{|\braket{n|\Phi}|}|\braket{n|\Psi}|\,,
\end{equation}
where \(\ket{\Psi}\) corresponds to the target state and \(\ket{\Phi}\) is chosen to be the trainable Slater-determinant. The trial is now non-continuous when elements of \(\braket{n|\Phi}\) are passing though zero, which complicates the training procedure.

If the state is close to a Slater-determinant, as indicated by a small value for \(S(\gamma_1)\), one already trivially expects a high fidelity for this approach, as the whole wavefunction including the amplitude can be recovered approximately by only considering Slater-determinants. A Slater-determinant can only be insufficient to determine the phase-structure if \(S(\gamma_1)\) is large enough. In other words, with this approach we probe how strongly the sign structure of the ground-state wave-function deviates from the one of the slater-determinant of the single particle orbitals. 

\begin{figure}[!hbt]
    \centering
    \includegraphics[width=\linewidth]{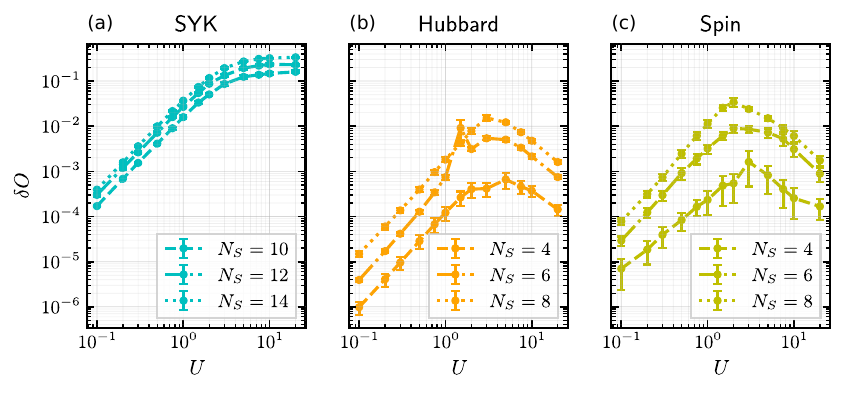}
    \caption{Achieved infidelity when using the phase of a Slater-determinant and the amplitude of the exact wavefunction for the groundstate of the SYK-model (a), the Hubbard model (b) and the model featuring a spin interaction (c) for various system sizes and interaction strengths \(U\) and \(J\) fixed at one. The results were averaged over ten realizations.  We observe that a Slater determinant not able to capture the sign structure of the SYK-model in the large interaction limit, while the performance on the other models is far better throughout the whole parameter range.}
    \label{fig:sign_SYK}
\end{figure}

\begin{figure}
    \centering
    \includegraphics[width=0.7\linewidth]{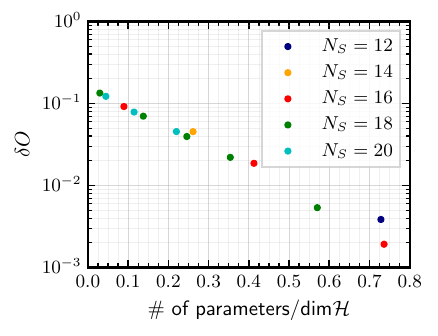}
    \caption{Here we have used a real multilayer perceptron with selu-hidden activation functions and \(4 N_S\) neurons per hidden layer and various depths for the logarithm of the amplitude of the wavefunction, while the phase was fixed by the exact phase of the groundstate obtained via exact diagonalization. No error-bars are shown, because the errors are smaller than the markers used in all cases. The relative number of hidden parameters required to achieve a given accuracy is again approximately proportional to the dimension of the Hilbert space indicating exponential scaling of the parameters required in the system size. The problems of NQSs when applied to the groundstates of the SYK-model can not be explained by the inability of neural quantum states to learn the correct sign structure.}\label{fig:syk_amplitude}
\end{figure}

Results for the SYK-model can be found in Fig.~\ref{fig:sign_SYK} (a). For this model, we observe that, while able to achieve a high fidelity in the weak interaction limit, a Slater-determinant is insufficient to capture the complex phase structure of the SYK-model in the strongly interacting limit. Yet, even when the proper sign-structure is restored, are still not able to represent the groundstate of the SYK-model in the absence of a kinetic term (see Fig.~\ref{fig:syk_amplitude}).

Qualitatively different behavior can be observed in Fig.~\ref{fig:sign_SYK} (b-c). These plots show the results for the trial states defined in eq.~\ref{eq:phase} trained on the groundstate of the Hubbard-model and the model with the spin interaction respectively. For these Slater-determinants are not only able to capture the sign structure in the weak interaction limit but also in the strong interaction limit. This indicates a deviation from the behavior observed for the SYK-model. Nonetheless, it is not clear weather this is due to the ground state again obtaining a more Slater-determinant like sign-structure or due to the ability of the Slater-determinant to more easily over fit the phase structure of the most important subspace due to an effective reduction of dimension in this limit, see Appendix~\ref{appendix:CHD}.

\begin{figure}[!hbt]
    \centering
    \includegraphics[width=\linewidth]{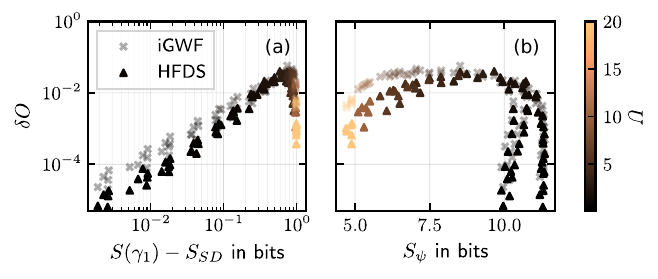}
    \caption{\(\delta O\) as a function of von-Neumann entroy  of the 1-RDM (a) and the entropy of the Born-distribution (b) for various interaction strengths for the fully connected Hubbard-Model with eight sites for \(M=1\) as well as for the inhomogenous Gutzwiller wavefunction.  Here no hidden layers were used. The $J$ controlling the kinetic term is fixed to one.}
    \label{fig:Hub-UiGW}
\end{figure}

\begin{figure}[!hbt]
    \centering
    \includegraphics[width=\linewidth]{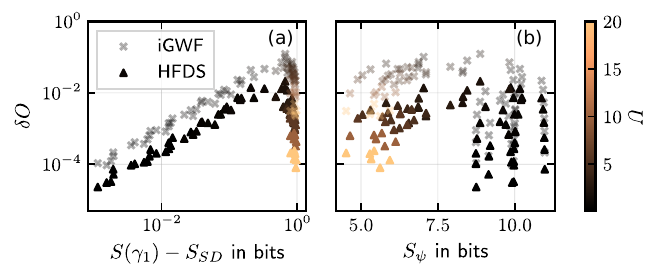}
    \caption{\(\delta O\) as a function of von-Neumann entroy  of the 1-RDM (a) and the entropy of the Born-distribution (b) for various interaction strengths for the model with the spin interaction channel with eight sites for \(M=1\) as well as for the inhomogenous Gutzwiller wavefunction.  Here no hidden layers were used. The $J$ controlling the kinetic term is fixed to one.}
    \label{fig:SPIN-UiGW}
\end{figure}

If the phase-structure of the groundstate is approximately given by a Slater-determinant, a Slater-determinant with a Gutzwiller style correlator is a reasonable variational ansatz. This ansatz is able to make use of both these properties when constructing approximations for the groundstate. In the following we compare the HFSD ansatz to the an inhomogeneous Gutziller correlator acting on a Slater-determinant:
\begin{equation}
    \braket{n|\Psi_{iGWF}}\propto \braket{n|\mathrm{e}^{-\sum_i \lambda_i \hat{n}_i^\uparrow\hat{n}_i^\downarrow}|\Phi}\,.
\end{equation}
Results for this wavefunction on the Hubbard model can be found in Fig.~\ref{fig:Hub-UiGW} and on the model featurig the spin-spin interaction in Fig.~\ref{fig:SPIN-UiGW}. The ability to achieve low infidelities on the groundstate indicates that the actual target state can be approximated reasonably well by a simple Gutzwiller-style wavefunction. This indicates that for both models the groundstate is close to a Gutzwiller-style wavefunction, at least for the small system sizes tested here, leading again to a reduction in complexity.

Remarkably, in case of the Hubbard model  the achieved error with the inhomogenous Gutzwiller wavefunction is already similar to the HFDS's error, even though it requires far fewer parameters.
Furthermore, we observed that the overlap only improves in the large interaction limit, once the the problems becomes effectively fragmented pointing again to this effect being due to the fragmentation and not due to the phase becoming slater-like.

\section{Entanglement structure of the Hubbard-model}\label{app:ent_structure}
\begin{figure}
    \centering
    \includegraphics[width=0.5\linewidth]{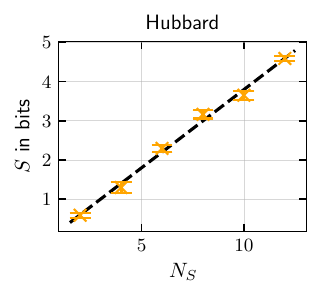}
    \caption{Entanglement for the Hubbard model when choosing the bipartition between the two spin sectors instead of splitting based on lattice sites for \(U=3\). This reduces the entanglement compared to what was shown in Fig.~\ref{fig:Entanglement}, indicating that its ground-state is expected to be far more compressible than the SYK-model.}\label{fig:entanglement_Hubbard}
\end{figure}
For the Hubbard model the entanglement within the spin subsectors is generated by the kinetic term. This entanglement can be captured due to the Slater-determinant based construction of the trial state used. In contrast the interaction term, the part for which makes the usage of the neural networks necessary, generates, to lowest order, local entanglement between the two spin sectors. This distinguishes the Hubbard-model from the remaining models, because for the remaining models also the interaction can generate non-local entanglement. For this reason have we also calculated the entanglement of the Hubbard model for a bipartition between the spin sectors. Considering this bipartition instead of a local bipartition leads to a reduction in entanglement as can be seen in Fig.~\ref{fig:entanglement_Hubbard} when compared to Fig.~\ref{fig:Entanglement}.

\begin{figure}
    \centering
    \includegraphics[width=\linewidth]{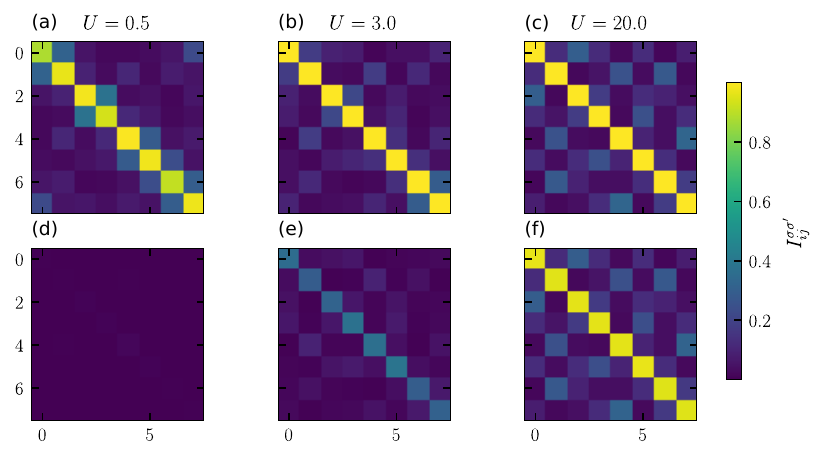}
    \caption{Mutual information for the fully connected Hubbard-model with \(N_S=8\) at half filling for various \(U\). The kinetic energy scale was fixed by setting \(J=1\). The upper row contains the results for \(\sigma=\sigma^\prime\) and the lower row the results for \(\sigma\neq\sigma^\prime\). The results are averaged over six realizations. Here the localization of the entanglement can clearly be seen.}\label{fig:mutual_information}
\end{figure}

Yet, parts of the entanglement shown in Fig.~\ref{fig:entanglement_Hubbard} are still not exponentially hard to capture, as for instance the suppression of double occupation in the large \(U\)-limit, an effect that could be described by a simple Gutzwiller-correlator as done in Fig.~\ref{fig:Hub-UiGW}, is generating entanglement. To further understand this, one can considering the mutual information
\begin{equation}
    I_{ij}^{\sigma \sigma^\prime}=S(\rho_i^\sigma)+S(\rho_j^\sigma)-S(\rho_{ij}^{\sigma\sigma^\prime})\,,
\end{equation}
where \(\rho_i^\sigma\) is the density matrix corresponding to the orbital \(i\) with spin \(\sigma\) and \(\rho_{ij}^{\sigma\sigma^\prime}\) is the density matrix corresponding to the orbitals \((i,\sigma)\) and \((j,\sigma^\prime)\) respectively. Here we have defined \(\rho_{ii}^{\sigma\sigma}:=\rho_i^\sigma\), see Fig.~\ref{fig:mutual_information} for the Hubbard model case. In subplots (b,e) we observe that the local Hubbard-style interaction are leading to local entanglement on the sites between the spin-sectors.

\begin{figure}
    \centering
    \includegraphics[width=\linewidth]{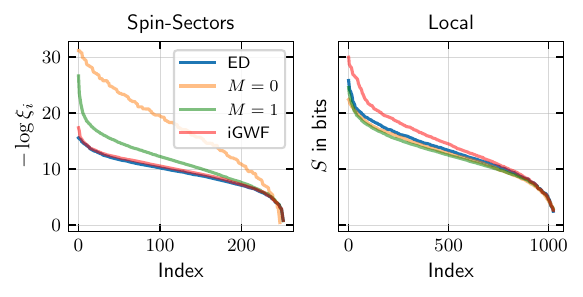}
    \caption{Entropy spectrum of the singlular values \(\xi_i\) of the reduced density matrix calculated for a realization of the fully connected Hubbard model with \(N_S=10\) and \(U/J=3\). Here the bipartition by spin sector is compared to a local bipartition. We compare a HFDS without hidden layers for \(M=1\) as well as \(M=1\)  to exact results as well as the inhomogenous Gutzwiller wavefunction. \(\xi_i<10^{-14}\) where discarded.}
    \label{fig:ent_spectrum}
\end{figure}

This general structure of the entanglement is further highlighted in Fig.~\ref{fig:ent_spectrum}. The spectrum corresponding to a local bipartition of the system can be captured by a slater-determinant, while a clear deviation from the exact results can be seen for the bipartition with respect to the spin sectors. However, the entanglement between the spin sectors is captured by a inhomogenous Gutzwiller wavefunction. Furthermore, HFDS are already able to improve significantly once a single hidden particle is considered.

\section{Away from half filling}\label{appendix:CHD}
\begin{figure}[!hbt]
    \centering
    \includegraphics[width=\linewidth]{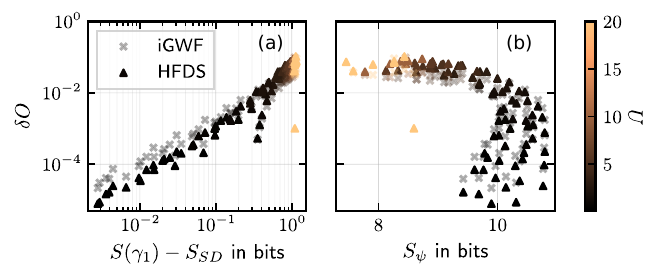}
    \caption{Same as Fig.~\ref{fig:Hub-UiGW}, but with \(3\) instead of \(4\) electrons per spin species. \(\delta O\) as a function of von-Neumann entroy  of the 1-RDM (a) and the entropy of the Born-distribution (b) for various interaction strengths for the fully connected Hubbard-Model with eight sites for HFDS with \(M=1\) as well as the inhomogeneous Gutzwiller wavefunction. Here again no hidden layers were used. The $J$ controlling the kinetic term is fixed to one.}
    \label{fig:HubD-U}
\end{figure}

\begin{figure}[!t]
    \centering
    \includegraphics[width=0.5\linewidth]{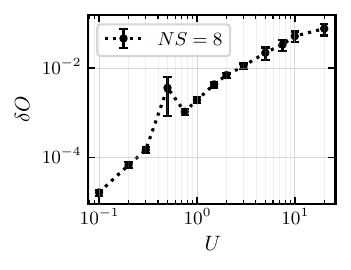}
    \caption{Achieved infidelity when using the phase of a Slater-determinant and the amplitude of the exact wavefunction trained on the groundstate of the Hubbard-model with \(3\) particles per spin species for various interaction strengths \(U\). \(J\) is fixed at one. The results were averaged over ten realizations. }
    \label{fig:sign_CHD}
\end{figure}

\begin{figure*}[!th]
    \includegraphics[width=0.6\linewidth]{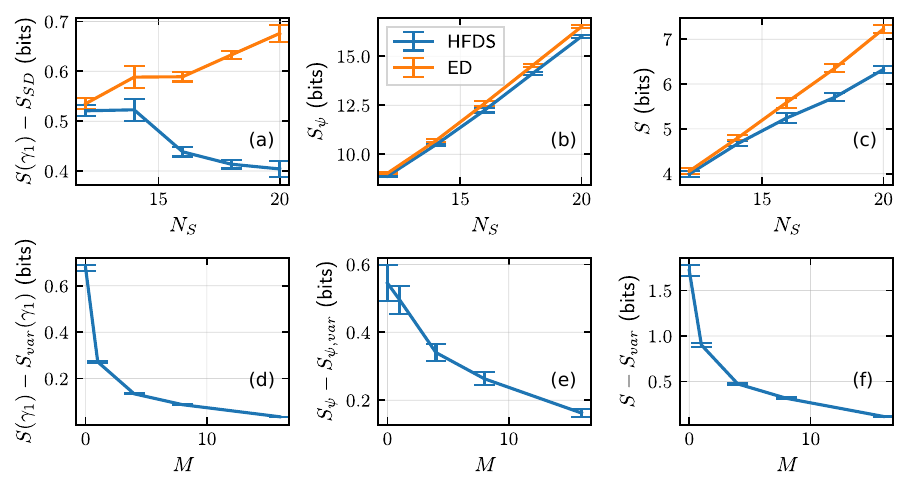}
    \caption{Comparison of the complexity measures for the SYK-model for the HFDS with two hidden layers and the exact state. In the first row (a-c) are the results for \(M=1\) and various system sizes. In the second row (d-f) are the results for \(N_S=20\) and various \(M\). For the estimation of the means and errors we average over six realizations. The results for the entanglement-entropy of the \(1\)-RDM are found in the first column (a,d), for the Shannon-entropy of the born distribution in the second column (b,e) and for the bipartition entanglement entropy in the last column (c,f).}\label{fig:compression}
\end{figure*}

In this appendix, we consider the effect of changed filling on the ability to represent the wavefunction. For this purpose we consider the Hubbard model with one electron per spin species removed and compare these results to Fig.~\ref{fig:Hub-UiGW}. Naively one would expect worse performance in the large \(U\) limit as the ground-state sector is now larger. This is confirmed in our calculations, see Fig.~\ref{fig:HubD-U}.
Furthermore, this model features a phase-structure for which the Slater-determinant achieves worse performance in the large interaction limit as shown in Fig.~\ref{fig:sign_CHD} when compared to the case of half filling shown in Fig.~\ref{fig:sign_SYK} (b) also leading to worse performance of the inhomogenous Gutzwiller wavefunction in the large interaction limit as shown in Fig.~\ref{fig:sign_SYK} (b) when compared to the case of half filling. However, it is not clear whether this is an actual feature of the sign structure or simply due to the dimension of the ground state block being strictly separated from the Slater-determinants in terms of parameter number.

\section{Complexities and Compression}
The three complexity measures, entanglement-entropy, entropy of the Born-distribution and the von-Neumann-entropy of the \(1\)-RDM, considered above all allow an efficient compression of the ground-state. However, as shown in Fig.~\ref{fig:compression}, the ground-state is approximated by less correlated, less entangled states with lower entropy of the Born distribution.
Furthermore, in the second row of Fig.~\ref{fig:compression} we demonstrate that we are able to converge these differences to zero by increasing the number of hidden particles for the largest system size considered here.
In general, we find the rather simple statement that complex states are approximated by simpler ones.
Furthermore, while the exact compression performed by neural networks usually can not be understood, these measures give insights how to improve the networks. Especially in case of the entanglement entropy it would be interesting to investigate whether it is more localized than for the exact ground state.
Fig.~\ref{fig:compression} also shows the growths of the entanglement entropy of the trained trial state.

\end{document}